\documentclass[runningheads]{llncs}
\usepackage{graphicx}
\usepackage{tikz}
\usepackage{comment}
\usepackage{amsmath,amssymb} 
\usepackage{color, booktabs, multirow, subfigure}
\usepackage[figuresleft]{rotating}
\usepackage{adjustbox}
\usepackage[pagebackref,breaklinks,colorlinks]{hyperref}
\usepackage[accsupp]{axessibility}  

\begin{document}
\pagestyle{headings}
\mainmatter
\def\ECCVSubNumber{100}  

\title{AIM 2022 Challenge on \\ Super-Resolution of Compressed Image and Video: Dataset, Methods and Results} 

\titlerunning{AIM 2022 Challenge on Super-Resolution of Compressed Image and Video}
%
\author{
Ren Yang \and Radu Timofte \and
Xin Li \and Qi Zhang \and Lin Zhang \and Fanglong Liu \and Dongliang He \and Fu li \and He Zheng \and Weihang Yuan \and
Pavel Ostyakov \and Dmitry~Vyal \and Magauiya Zhussip \and Xueyi Zou \and Youliang Yan \and
Lei Li \and Jingzhu~Tang \and Ming~Chen \and Shijie Zhao \and
Yu Zhu \and Xiaoran Qin\and Chenghua Li\and Cong~Leng\and Jian~Cheng\and
Claudio Rota  \and Marco Buzzelli \and Simone Bianco \and Raimondo~Schettini \and
Dafeng Zhang\and Feiyu Huang\and Shizhuo Liu\and Xiaobing~Wang\and Zhezhu Jin\and
Bingchen Li\and Xin Li  \and
Mingxi Li\and Ding Liu \and
Wenbin~Zou \and Peijie Dong \and Tian Ye\and Yunchen Zhang \and Ming Tan \and Xin~Niu\and
Mustafa Ayazoglu\and
Marcos Conde\and Ui-Jin Choi\and
Zhuang Jia  \and Tianyu Xu\and Yijian~Zhang\and
Mao Ye  \and Dengyan Luo  \and Xiaofeng Pan  \and
Liuhan Peng
}
\authorrunning{Ren Yang, Radu Timofte et al.}
%
\institute{
Organizers: \email{ren.yang@vision.ee.ethz.ch, radu.timofte@uni-wuerzburg.de}}
\maketitle

\begin{abstract}
 This paper reviews the Challenge on Super-Resolution of Compressed Image and Video at AIM 2022. This challenge includes two tracks. Track 1 aims at the super-resolution of compressed image, and Track~2 targets the super-resolution of compressed video. In Track 1, we use the popular dataset DIV2K as the training, validation and test sets. In Track 2, we propose the LDV 3.0 dataset, which contains 365 videos, including the LDV 2.0 dataset (335 videos) and 30 additional videos. In this challenge, there are 12 teams and 2 teams that submitted the final results to Track 1 and Track 2, respectively. The proposed methods and solutions gauge the state-of-the-art of super-resolution on compressed image and video. The proposed LDV 3.0 dataset is available at \url{https://github.com/RenYang-home/LDV\_dataset}. The homepage of this challenge is at \url{https://github.com/RenYang-home/AIM22\_CompressSR}.
\end{abstract}

\section{Introduction}

Compression plays an important role on the efficient transmission of images and videos through the band-limited Internet. However, image and video compression unavoidably leads to compression artifacts, which may severely degrade the visual quality. Therefore, quality enhancement of compressed image and video has become a popular research topic. However, in the early years, due to the limitation of devices and band-width, the image and video are usually with low resolution. Therefore, when we intend to restore them to high resolution and good quality, we face the challenge to achieve both super-resolution and quality enhancement of compressed image (Track 1) and video (Track 2).

In the past decade, a great number of works were proposed for single image super-resolution~\cite{dong2014learning,kim2016accurate,sajjadi2017enhancenet,zhang2017beyond,zhang2018image,zhang2018residual,deng2019wavelet,liang2021swinir} and there are also plenty of methods proposed for the reduction of JPEG artifacts~\cite{dong2015compression,zhang2017beyond,tai2017memnet,ehrlich2020quantization,jiang2021towards}. Recently, the blind super-resolution~\cite{gu2019blind,yamac2021kernelnet,zhang2021designing} methods have been proposed. They are able to use one model to jointly handle the tasks of super-resolution, debluring, JPEG artifacts reduction, etc. Meanwhile, video super-solution~\cite{caballero2017real,tao2017detail,wang2019edvr,isobe2020video,chan2021basicvsr,chan2021basicvsr++,liang2022vrt,liu2022video} and compression artifacts reduction~\cite{yang2017decoder,yang2018enhancing,yang2018multi,guan2019mfqe,yang2019quality,Xu_2019_ICCV,yang2020learning} also has become a popular topic, which aims at adequately exploring the temporal correlation among frames to facilitate the super-resolution and quality enhancement of videos. NTIRE 2022~\cite{yang2022ntire} is the first 
challenge we organized on super-resolution of compressed video. The winner method~\cite{zheng2022progressive} in the NTIRE 2022 challenge successfully outperforms the state-of-the-art method~\cite{chen2021compressed}.

The AIM 2022 Challenge on Super-Resolution of Compressed Image and Video is one of the AIM 2022 associated challenges: reversed ISP~\cite{conde2022aim}, efficient learned ISP~\cite{ignatov2022isp}, super-resolution of compressed image and video~\cite{yang2022aim}, efficient image super-resolution~\cite{ignatov2022isr}, efficient video super-resolution~\cite{ignatov2022vsr}, efficient Bokeh effect rendering~\cite{ignatov2022bokeh}, efficient monocular depth estimation~\cite{ignatov2022depth}, Instagram filter removal~\cite{kinli2022aim}.

The AIM 2022 Challenge on Super-Resolution of Compressed Image and Video steps forward for establishing a benchmark of the super-resolution of JPEG image (Track 1) and HEVC video (Track 2). The methods proposed in this challenge are also have the potential to solve various super-resolution tasks. In this challenge, Track 1 utilizes the DIV2K~\cite{agustsson2017ntire} dataset, and Track 2 uses the proposed LDV 3.0 dataset, which contains 365 videos with diverse content, motion, frame-rate, etc. In the following, we first describe the AIM 2022 Challenge, including the DIV2K~\cite{agustsson2017ntire} dataset and the proposed LDV 3.0 dataset. Then, we introduce the proposed methods and the results.

\section{AIM 2022 Challenge}

The objectives of the AIM 2022 challenge on Super-Resolution of Compressed Image and Video are: (i) to advance the state-of-the-art in super-resolution of compressed inputs; (ii) to compare different solutions; (iii) to promote the proposed LDV 3.0 dataset.

\subsection{DIV2K~\cite{agustsson2017ntire} dataset}

The DIV2K~\cite{agustsson2017ntire} dataset consists of 1,000 high-resolution images with diverse contents. In Track 1 of AIM 2022 Challenge, we use the training (800 images), validation (100 images) and test (100 images) sets of DIV2K for training, validation and test, respectively.

\subsection{LDV 3.0 dataset}

The proposed LDV 3.0 dataset is an extension of the LDV 2.0 dataset~\cite{yang2022ntire,yang2021ntire,yang2021dataset} with 30 additional videos. Therefore, there are totally 365 videos in the LDV~3.0 dataset. The same as LDV and LDV 2.0, the additional videos in LDV 3.0 are collected from YouTube~\cite{youtube}, containing 10 categories of scenes, i.e., \textit{animal}, \textit{city}, \textit{close-up}, \textit{fashion}, \textit{human}, \textit{indoor}, \textit{park}, \textit{scenery}, \textit{sports} and \textit{vehicle}, and they are with diverse frame-rates from 24 fps to 60 fps. To ensure the high quality of the groundtruth videos, we only collect the videos with 4K resolution, and without obvious compression artifacts. We downscale the videos to further remove the artifacts, and crop the width and height of each video to the multiples of 8, due to the requirement of the HEVC test model (HM). Besides, we convert videos to the format of YUV 4:2:0, which is the most commonly used format in the existing literature. Note that all source videos in our LDV 3.0 dataset have the licence of \emph{Creative Commons Attribution licence (reuse allowed)}\footnote{\url{https://support.google.com/youtube/answer/2797468?hl=en}}, and our LDV~3.0 dataset is used for academic and research proposes.

The Track 2 of AIM 2022 Challenge has the same task as the Track 3 of our NTIRE 2022 Challenge~\cite{yang2022ntire}. Therefore, we use the training, validation and test sets of the Track 3 in NTIRE 2022 as the training set (totally 270 videos) for the Track 2 in AIM 2022 Challenge. All videos in the proposed LDV, LDV 2.0 and LDV 3.0 datasets and the splits in NTIRE 2021, NTIRE 2022 and AIM 2022 Challenges are publicly available at \url{https://github.com/RenYang-home/LDV_dataset}.

\subsection{Track 1 -- super-resolution of compressed image}

JPEG is the most commonly used image compression standard. Track 1 targets the $\times 4$ super-resolution of the images compressed with JPEG with the quality factor of 10. Specifically, we use the following Python codes to produce the low resolution samples:

\ 

\noindent\texttt{from PIL import Image}

\noindent\texttt{img = Image.open(path\_gt + str(i).zfill(4) + '.png')}

\noindent\texttt{w, h = img.size}

\noindent\texttt{assert w \% 4 == 0} \\
\noindent\texttt{assert h \% 4 == 0}

\noindent\texttt{img = img.resize((int(w/4), int(h/4)), resample=Image.BICUBIC)}

\noindent\texttt{img.save(path + str(i).zfill(4) + '.jpg', "JPEG", quality=10)}

\ 

\noindent In this challenge, we the version 7.2.0 of the Pillow library.

\subsection{Track 2 -- super-resolution of compressed video}

Track 2 has the same task as the Track 3 in NTIRE 2022~\cite{yang2022ntire}, which requires the participants to enhance and meanwhile $\times 4$ super-resolve the HEVC compressed video. In this track, the input videos are first downsampled by the following command:

\ 

\noindent\texttt{ffmpeg -pix\_fmt yuv420p -s WxH -i x.yuv \\ -vf scale=(W/4)x(H/4):flags=bicubic x\_down.yuv}

\ 

\noindent where \texttt{x}, \texttt{W} and \texttt{H} indicates the video name, width and height, respectively. Then, the downsampled video is compressed by HM 16.20\footnote{\url{https://hevc.hhi.fraunhofer.de/svn/svn_HEVCSoftware/tags/HM-16.20}} at QP = 37 with the default Low-Delay P (LDP) setting (\textit{encoder\_lowdelay\_P\_main.cfg}). Note that, we first crop the groundtruth videos to make sure that the downsampled width (\texttt{W/4}) and height (\texttt{H/4}) are integer numbers.

\section{Challenge results}

\subsection{Track 1}

The PSNR results and the running time of Track 1 are shown in Table~\ref{tab:1}. In this track, we use the images that are directly upscaled by the bicubic algorithm as the baseline. As we can see from Table~\ref{tab:1}, all methods proposed in this challenge achieves $>1$ dB PSNR improvement over the baseline. The PSNR improvement of the top 3 methods are higher than 1.3 dB over the baseline. The VUE Team achieves the best result, that is $\sim 0.1$ dB higher the runner-up method. We can also see from Table~\ref{tab:1} that the top methods consume high time complexity, while the method of the Giantpandacv Team is the most time-efficient one, whose running time is significantly lower than the methods with higher PSNR. Note that, the data in Table~\ref{tab:1} are provided by the participants, so the data may be obtained under different hardware and conditions. Therefore, Table~\ref{tab:1} is only for reference. It is hard to guarantee the fairness in comparing time efficiency.

The test and training details are presented in Table~\ref{tab:2}. As Table~\ref{tab:2} shows, most methods use extra training data to improve the performance. In this challenge, Flickr2K~\cite{timofte2017ntire} is the most popular dataset used in training, in addition to the official training data provided by the organizers. In inference, the self-ensemble strategy~\cite{timofte2016seven} is widely utilized. It has been proved to be an effective skill to boost the performance of super-resolution.

\begin{table}[!t]
\setlength\tabcolsep{4pt}
\scriptsize
  \centering
  \caption{Results of Track 1 ($\times 4$ super-resolution of JPEG image). The test input is available at \url{https://codalab.lisn.upsaclay.fr/competitions/5076}, and the researchers can submit their results to the ``testing'' phase at the CodaLab server to get the performance of their methods to compare with the numbers in this table. 
  }\label{tab:1}
    \begin{tabular}{cccc}
    \toprule
    Team & PSNR (dB) & Running time (s) & Hardware  \\
    \midrule
    VUE &23.6677  &120 &Tesla V100 \\
BSR~\cite{li2022multi-patch} & 23.5731& 63.96 & Tesla A100\\
CASIA LCVG~\cite{qin2022cidbnet} & 23.5597& 78.09 &Tesla A100 \\
SRC-B &23.5307 & 18.61 &GeForce RTX 3090 \\
USTC-IR~\cite{li2022hst} & 23.5085& 19.2 & GeForce 2080ti \\
MSDRSR & 23.4545& 7.94 &Tesla V100 \\
Giantpandacv & 23.4249& 0.248 & GeForce RTX 3090 \\
Aselsan Research & 23.4239& 1.5 &GeForce RTX 2080\\
SRMUI~\cite{conde2022swin2sr} & 23.4033 & 9.39& Tesla A100 \\
MVideo & 23.3250 & 1.7 & GeForce RTX 3090 \\
UESTC+XJU CV & 23.2911& 3.0 & GeForce RTX 3090 \\
cvlab & 23.2828 & 6.0 & GeForce 1080 Ti \\
\midrule
Bicubic $\times 4$ &22.2420 & - & -\\
\toprule
    \end{tabular}%
\end{table}%

\begin{table}
\setlength\tabcolsep{4pt}
\scriptsize
  \centering
  \caption{Test and training details of Track 1 ($\times 4$ super-resolution of JPEG image). \label{tab:2}
  }
    \begin{tabular}{ccc}
    \toprule
    Team &  Ensemble for test  & Extra training data \\
    \midrule
    VUE  &Flip/rotation & ImageNet \cite{deng2009imagenet}, Flickr2K~\cite{timofte2017ntire}\\
BSR &Flip/rotation, three models for voting & Flickr2K~\cite{timofte2017ntire}, Flickr2K-L$^3$\\
CASIA LCVG & Flip/rotation, three models& ImageNet \cite{deng2009imagenet}\\
SRC-B  & Flip/rotation& Flickr2K~\cite{timofte2017ntire}\\
USTC-IR & Flip/rotation & Flickr2K~\cite{timofte2017ntire} and CLIC datasets \\
MSDRSR  & Flip/rotation& Flickr2K~\cite{timofte2017ntire} and DIV8K~\cite{gu2019div8k}\\
Giantpandacv & Flip/rotation, TLC~\cite{TLC} & Flickr2K~\cite{timofte2017ntire}\\
Aselsan Research & Flip/rotation& Flickr2K~\cite{timofte2017ntire}\\
SRMUI & Flip/rotation & Flickr2K~\cite{timofte2017ntire} and MIT 5K~\cite{fivek}\\
MVideo & Flip/rotation & -\\
UESTC+XJU CV  & Flip/rotation& - \\
cvlab & - & 4,600 images\\
\midrule
Bicubic $\times 4$  & - & -\\
\toprule
\multicolumn{3}{l}{$^3$Flicker2K-L is available as ``flickr2k-L.csv''at \url{https://github.com/RenYang-home/AIM22\_CompressSR/}} \\
    \end{tabular}%
\end{table}%

\subsection{Track 2}

Table~\ref{tab:3} shows the results of Track 2. Similar to Track 1, we use the videos that are directly upscaled by the bicubic algorithm as the baseline performance in this track. The winner team NoahTerminalCV improves the PSNR by more than 2 dB over the baseline, and it successfully beats the winner method~\cite{zheng2022progressive} in NTIRE 2022, which can be seen as the state-of-the-art method. The IVL method has the very fast running speed, and it is able to achieve the real-time super-resolution on the test videos. In Table~\ref{tab:4}, we can see that the NoahTerminalCV Team uses a large training set, including 90,000 videos collected from YouTube~\cite{youtube}. This may be obviously beneficial for their test performance. Note that, the data in Table~\ref{tab:4} are provided by the participants. It is hard to guarantee the fairness in comparing time efficiency.

\begin{table}[!t]
\scriptsize
\setlength\tabcolsep{4pt}
  \centering
  \caption{Results of Track 2 ($\times 4$ super-resolution of HEVC video). \textcolor{blue}{Blue} indicates the state-of-the-art method. The test input and groundtruth are available on the homepage (see the abstract) of the challenge.
  }
  \label{tab:3}
    \begin{tabular}{cccc}
    \toprule
    Team & PSNR (dB) & Time (s) & Hardware\\
    \midrule
    NoahTerminalCV & 25.1723 & 10 & Tesla V100 \\
\textcolor{blue}{NTIRE'22 Winner}~\cite{zheng2022progressive} & 24.1097& 13.0 & Tesla V100 \\
IVL & 23.0892 & 0.008 &GeForce GTX 1080  \\
\midrule
Bicubic $\times 4$ & 22.7926& - & - \\
\toprule

    \end{tabular}%
 
\end{table}

\begin{table}[!t]
\scriptsize
\setlength\tabcolsep{4pt}
  \centering
  \caption{Test and training details of Track 2 ($\times 4$ super-resolution of HEVC video). \textcolor{blue}{Blue} indicates the state-of-the-art method.
  }\label{tab:4}
    \begin{tabular}{cccccc}
    \toprule
    Team  & Ensemble for test & Extra training data \\
    \midrule
    NoahTerminalCV & Flip/rotation & 90K videos$^4$ from YouTube~\cite{youtube} \\
\textcolor{blue}{NTIRE'22 Winner}~\cite{zheng2022progressive} & Flip/rotation, two models & 870 videos from YouTube~\cite{youtube}\\
IVL &  - & - \\
\midrule
Bicubic $\times 4$ & - & - \\
\toprule

\multicolumn{3}{l}{$^4$Dataset is available as ``dataset\_Noah.txt'' at \url{https://github.com/RenYang-home/AIM22\_CompressSR/}} \\
    \end{tabular}%
 
\end{table}

\section{Teams and methods}
\subsection{VUE Team}

\begin{figure}[b]
	\begin{center}
		\includegraphics[width=\columnwidth]{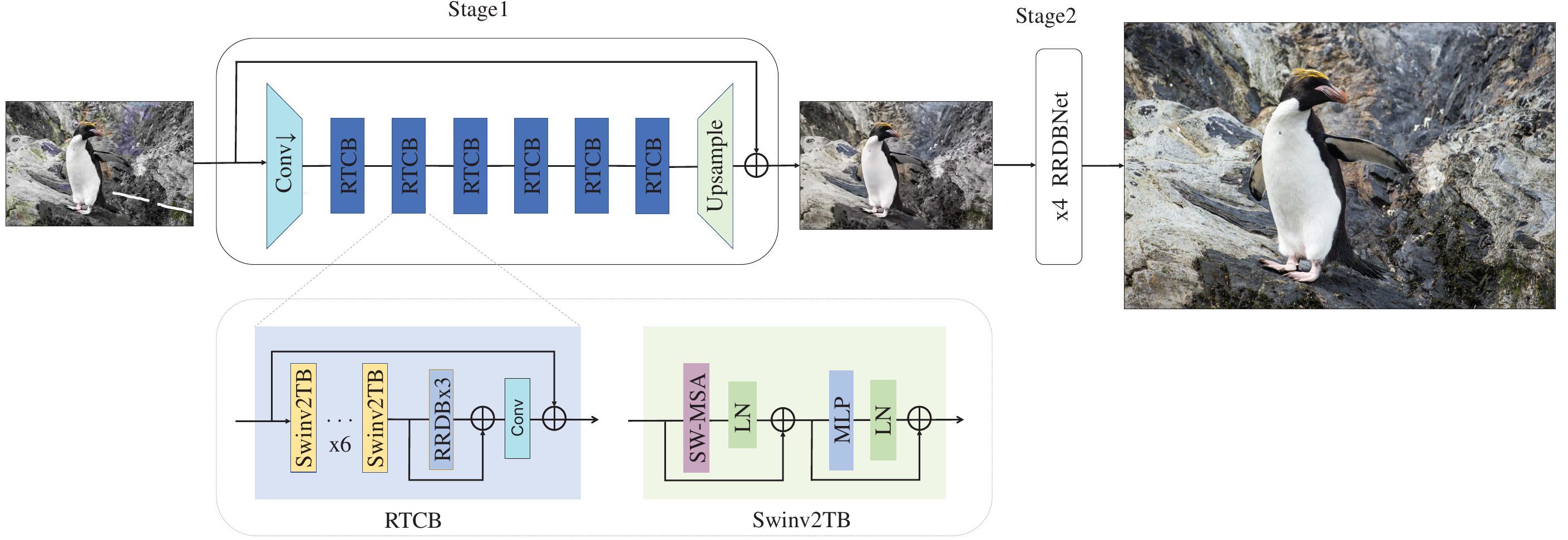}
	\end{center}
	\caption{Overview of the TCIR method proposed by the VUE Team.}
	\label{fig:Method}
\end{figure}

The method proposed by the VUE Team is called TCIR: A Transformer and CNN Hybrid Network for Image Restoration. The architecture of TCIR is shown in Fig.~\ref{fig:Method}. Specifically, they decouple this task of Track 1 into two sub-stages. In the first stage, they propose a Transformer and CNN hybrid network (TCIR) to remove JPEG artifacts, and in the second stage they use a finetuned RRDBNet for $\times$4 super-resolution. The proposed TCIR is based on SwinIR~\cite{liang2021swinir} and the main improvements are as follows:
\begin{itemize}
\item 1) They conduct $\times$2 downsampling to the JPEG-compressed input by a convolution with the stride of 2. The main purpose of this downsampling is to save GPU memory and speed up the model. Since the images compressed by JPEG with the quality factor of 10 are very blurry, this does not affect the performance of TCIR.
\item 2) Then, they use the new Swinv2 transformer block layer to replace the STL in the original SwinIR to greatly improve the capability of the network.
\item 3) In addition, they add several RRDB~\cite{zhang2018residual} modules to the basic blocks of TCIR and this combines the advantages of Transformer and CNN.
\end{itemize}

\subsection{NoahTerminalCV Team}

\begin{figure*}[t]
    \centering
    \includegraphics[width=.8\linewidth]{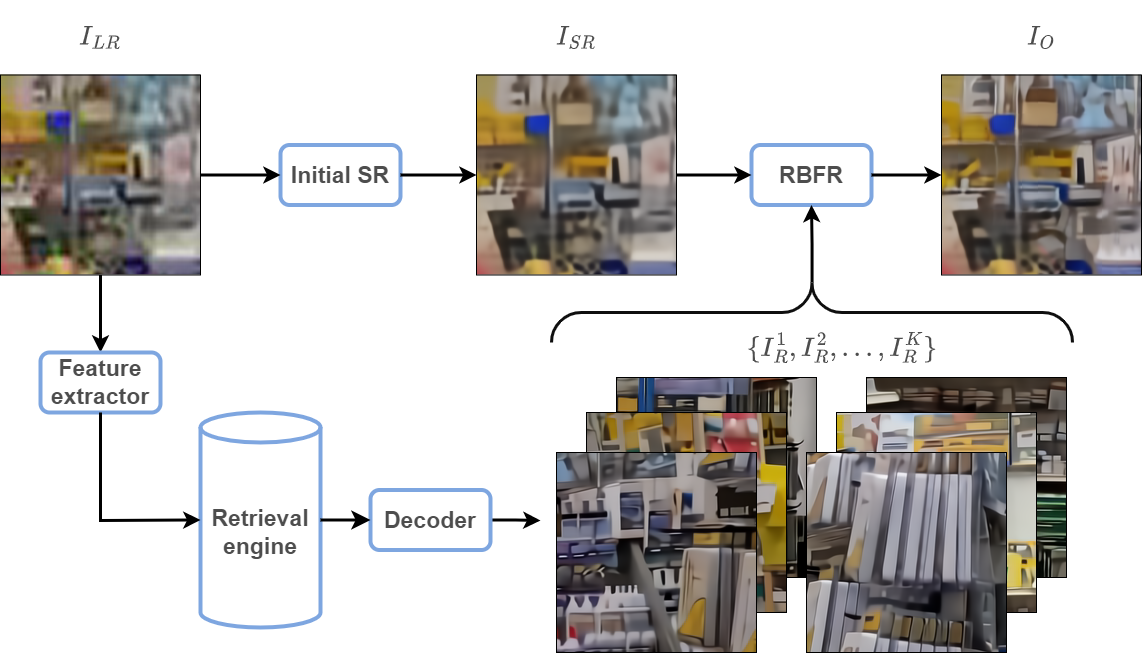}
      \caption{The method proposed by the NoahTerminalCV Team.}
  \label{fig:rbfr}
\end{figure*}

The method proposed by the NoahTerminalCV Team is called Enhanced Video Super-Resolution through Reference-Based Frame Refinement. As Fig.~\ref{fig:rbfr} shows, the proposed method consists of two subsequent stages. Firstly, they perform an initial super-resolution using a feed-forward multi-frame neural network. Then, the second step is called reference-based frame refinement. They find top K similar images for each low-resolution input frame from the external database. Then, they run a matching correction step for every patch on this input frame to perform a global alignment of reference patches. As a result, the $I_{SR}$, which comes from the first stage, and a set of globally aligned references $\{I_{R}^1, I_{R}^2,..., I_{R}^K\}$ are obtained. Finally, they are processed with RBFR network ($N_{RBFR}$) to handle residual misalignments and to properly transfer texture, details from reference images to initially super-resolved output $I_{SR}$. The details of training and test are described in the following.

\vspace{1em}
\noindent\textbf{4.2.1 Training}
\vspace{1em}

\textbf{Initial Super-Resolution (Initial SR).}
The NoahTerminalCV Team upgraded the BasicVSR++~\cite{chan2021basicvsr++} by increasing channels to 128 and reconstruction blocks to 45.
The BasicVSR++ is trained from scratch (except SPyNet) using a pixel-wise $L_1$ objective on the full input images without cropping and fine-tuned using $L_{1} + L_{2}$ objectives. The training phase took about 21 days using 8 NVIDIA Tesla V100 GPUs. They observed a slight performance boost if the model is fine-tuned with a combination of $L_{1}$ and $L_{2}$ losses.

\textbf{Reference-based Frame Refinement (RBFR).}
The information from subsequent frames is not always enough to produce a high-quality super-resolved image. Therefore, after initial super-resolution using upgraded BasicVSR++, they employed a reference-based refinement strategy. The idea is to design a retrieval engine that will find top K closest features in the database and then transfer details/texture from them to the initially upscaled frame. The retrieval engine includes feature extractor, database of features, and autoencoder.

\textbf{Feature Extractor.}
They trained a feature extractor network that takes a low-resolution image $I_{LR}$ and represents it as a feature vector. They used a contrastive learning~\cite{chen2020simple} framework to train the feature extractor: for positive samples, they used two random frames from the same video, while for the negative samples we employ frames from different videos. The backbone for a feature extractor was Resnet-34~\cite{he2016deep}.

\textbf{Database and Autoencoder.}
The database consists of 2,000,000 samples generated from the training dataset. Each sample is compressed into a feature $z = N_E({I_{HR}})$ using the Encoder $N_E$, since saving as images in a naive way is not practically plausible. Once they find top K similar features, the Decoder $N_D$ is used to reconstruct the original input $\widehat{I}_{HR} = N_D(z)$.

\textbf{Retrieval Engine.}
After compressing the database of images using the trained Encoder, obtaining latent representations, and representing all low-resolution versions as a feature vector of size 100 extracted from the trained Feature Extractor, they build an index using the HNSW~\cite{DBLP:journals/corr/MalkovY16} algorithm from the nmslib~\cite{DBLP:conf/sisap/BoytsovN13} library. This algorithm allows searching for the top K nearest neighbors in the database.

\textbf{RBFR.}
Finally, we train a network $N_{RBFR}$ that takes the result of initial super-resolution $I_{SR}$ and top K similar images from the database $\{I_{R}^1,I_{R}^2,...,I_{R}^K\}$. The network produces the final prediction $I_O$. We train $N_{RBFR}$ through the L1 objective between $I_O$ and $I_{HR}$. As a $N_{RBFR}$, we use NoahBurstSRNet~\cite{bhat2021ntire} architecture, since it effectively handles small misalignments and can properly transfer information from reference non-aligned images.

\vspace{1em}
\noindent\textbf{4.2.2 Test}
\vspace{1em}

\textbf{Initial Super-Resolution (Initial SR).}
During the inference, in order to upscale the key-frame $I_{LR}^i$, they put it to the initial super-resolution network together with additional frames $I_{LR}^{i-1}, I_{LR}^{i+1}, I_{LR}^{i-2}, I_{LR}^{i+2}...$. The number of additional frames during the inference is set to the full sequence size (up to 600 images).

\textbf{Reference-based Frame Refinement (RBFR).}
For RBFR, top K (typically 16) similar images are first obtained using the retrieval engine. Then, the inference is done in a patch-wise manner. They extract a patch from the $I_{SR}$ and use the Template Matching~\cite{briechle2001template} to perform a global alignment and find the most similar patches on the images $\{I_{R}^1, I_{R}^2,..., I_{R}^K\}$. Then, they put them to the $N_{RBFR}$ to generate the final result.

\subsection{BSR Team~\cite{li2022multi-patch}}

For most low-level tasks, like image super-resolution, the network is trained on cropped patches rather than the full image. That means the network can only look at the pixels inside the patches during the training phase, even though the network's ability is becoming more and more powerful and the receptive field of the deep neural network could be very large.

The patch size heavily affects the ability of the network. However, with the limitation of the memory and the computing power of GPU, it is not a sanity choose to train the network on the full image.
To address the above-mentioned problem, the BSR Team proposes a multi-patches method to greatly increase the receptive field in the training phase while increasing very little memory.

\begin{figure}
    \begin{center}
       \includegraphics[width=0.7\linewidth]{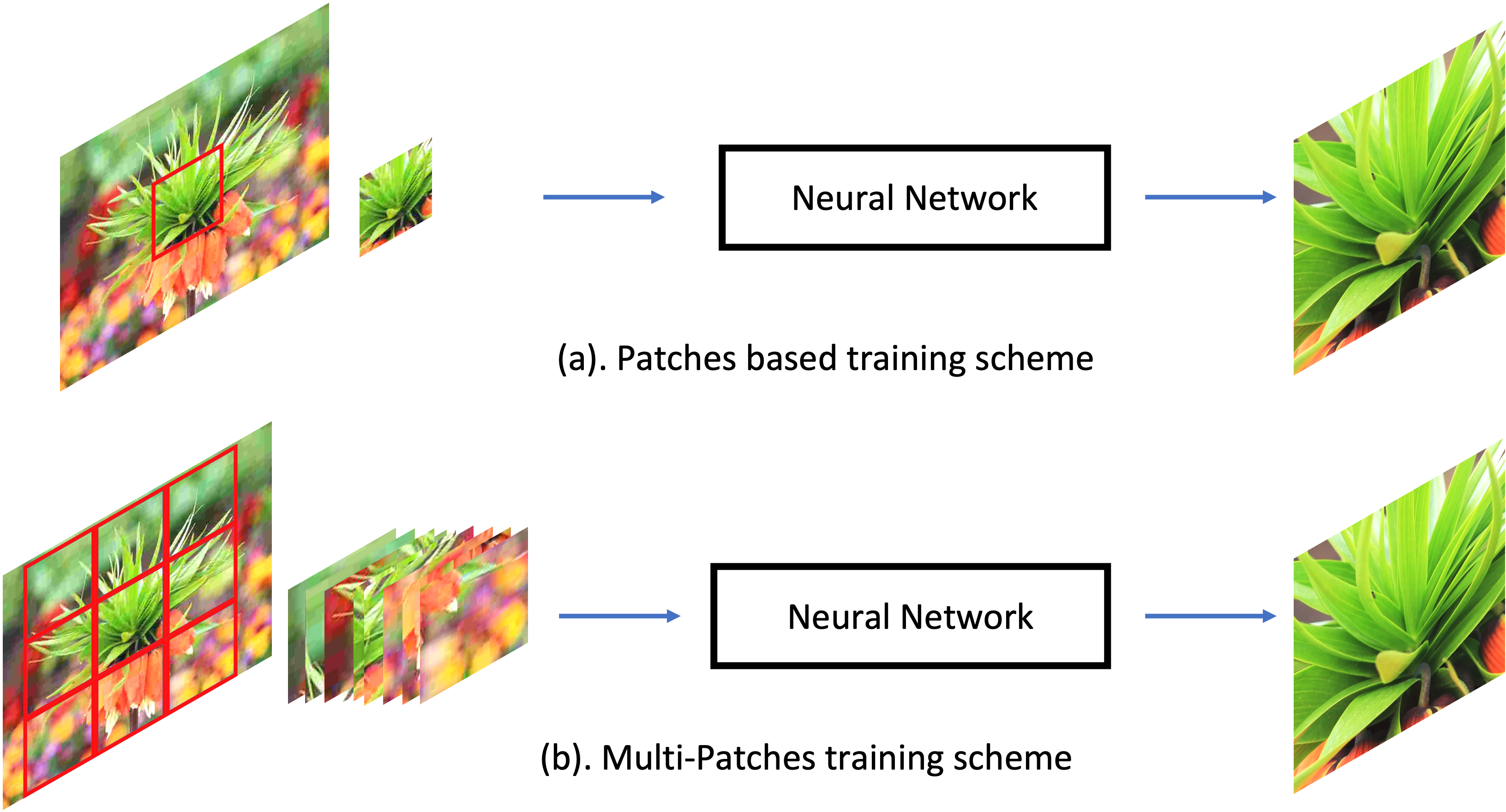}
    \end{center}
    \vspace{-1em}
       \caption{Illustration of the multi patches scheme proposed by the BSR Team. Top images represent general patches based training scheme and bottom image represent multi-patches scheme. Low resolution input patch and its eight surrounding patches are cropped then send to the neural network to reconstruct the super resolution image of the centre patch. The neural network chosen in this competition is HAT~\cite{chen2022activating}.}
    \label{fig:multi-patch-scheme}
\end{figure}

As shown in Fig.~\ref{fig:multi-patch-scheme}, they crop low-resolution input patch and its eight surrounding patches as our multi-patches network's input. They use HAT~\cite{chen2022activating} as the backbone, and propose Multi Patches Hybrid Attention Transformer (MPHAT). Compared with HAT~\cite{chen2022activating}, MPHAT just simply changew the input channel of the network for the multi-patch input. On the validation set of the challenge, the proposed MPHAT achieves the PSNR performance of 23.6266 dB, which is obviously higher than the HAT without multi-patches (23.2854 dB).

In the training phase, they train the network by using Adam optimizer with $\beta_1 = 0.9$ and $\beta_2 = 0.99$ to minimize the MSE loss. The model is trained for 800,000 iterations with mini-batches of size 32 and patch size 64. The learning rate is initialized as $10^{-4}$ and reduced to half at the 300,000th, 500,000th, 650,000th, 700,000th, 750,000th iterations, respectively. 

\subsection{CASIA LCVG Team~\cite{qin2022cidbnet}}

\begin{figure}[t]
  \begin{center}
     \includegraphics[scale=0.3]{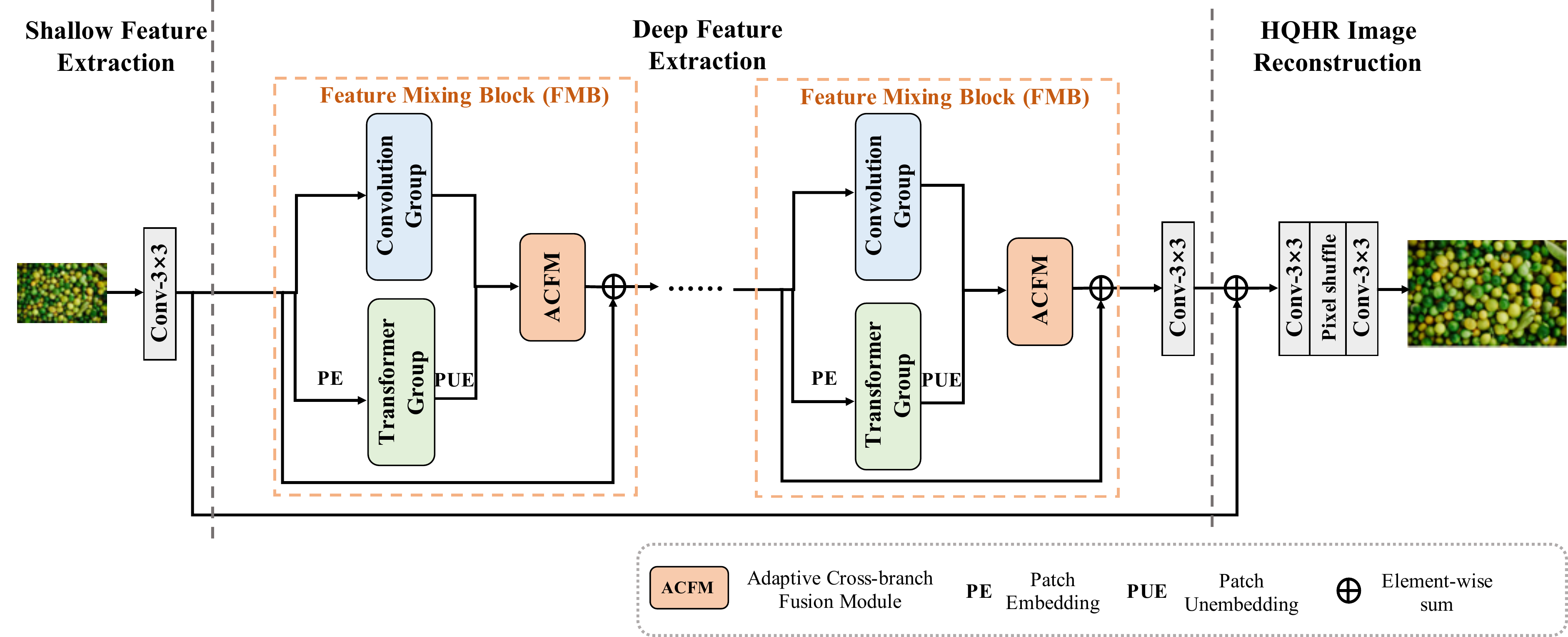}
      \caption{The overall architecture of the Consecutively-Interactive Dual-Branch network (CIDBNet) of the CASIA LCVG Team.}  
      \label{cidbnet} 
      \end{center}
\end{figure}

The CASIA LCVG Team proposes a consecutively-interactive dual-branch network (CIDBNet) to take advantage of both convolution and transformer operations, which are good at extracting local features and global interactions, respectively. To better aggregate the two-branch information, they newly introduce an adaptive cross-branch fusion module (ACFM), which adopts a cross-attention scheme to enhance the two-branch features and then fuses them weighted by a content-adaptive map. Experimental results demonstrate the effectiveness of CIDBNet, and in particular, CIDBNet achieves higher performance than a larger variant of HAT (HAT-L)~\cite{chen2022activating}. The framework of the proposed method is illustrated in Fig.~\ref{cidbnet}.

They adopt 1,280,000 images from ImageNet \cite{deng2009imagenet} as training set and all the models are trained from scratch. They set the input patch size to $64\times64$ and use random rotation and horizontally flipping for data augmentation. The mini-batch size is set to 32 and total training iterations are set to 800,000. The learning rate is initialized as $2\times 10^{-4}$. It remains constant for the first 270,000 iterations and then decreases to $10^{-6}$ in the next 560,000 iterations following the cosine annealing. They adopt the Adam optimizer $(\beta_{1}=0.9, \beta_{2}=0.9)$ to train the model. During test,  they first apply self-ensemble trick for each model, which could involves 8 outputs for fusion. Then, they fuse the self-ensembled outputs of the CIDBNet, CIDBNet\_NF and CIDBNet\_NFE models, respectively.

\subsection{IVL Team}

\begin{figure}
    \centering
    \includegraphics[width=.9\textwidth]{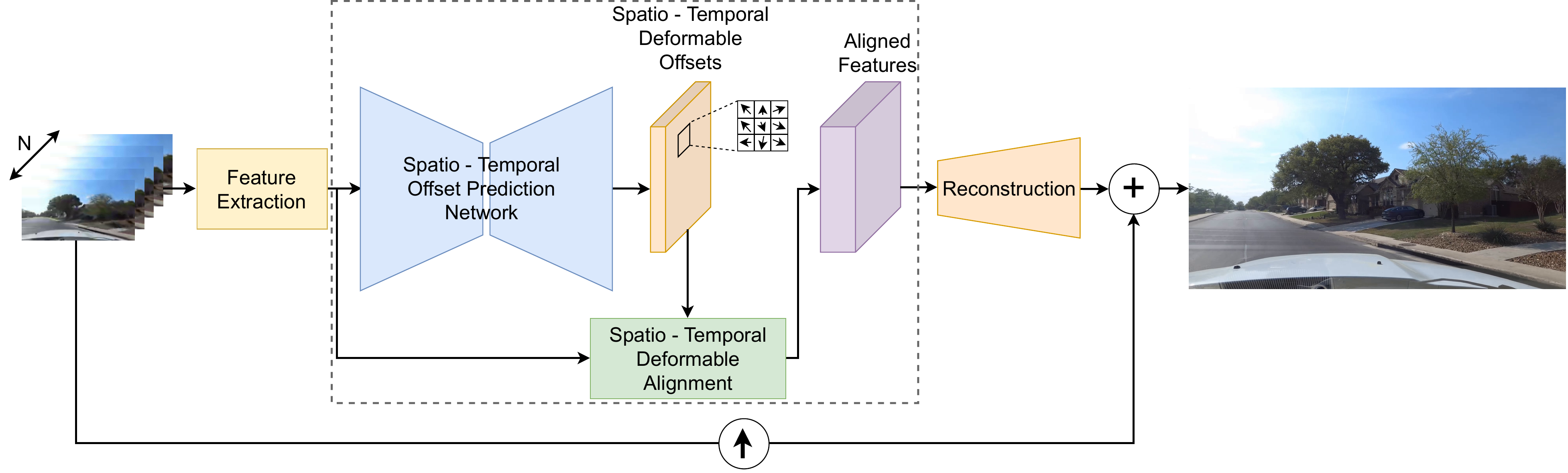}
    \caption{The method proposed by the IVL Team.}
    \label{fig:ivl_method}
\end{figure}

The architecture proposed by the IVL team for the video challenge track is shown in Fig.~\ref{fig:ivl_method} and contains three cascaded modules. The first module stacks the input frames (five consecutive frames are used) and extracts deep features from them. The second module aligns the features extracted from the adjacent frames with the features of the target frame. This is achieved by using a Spatio-Temporal Offset Prediction Network (STOPN), which implements a U-Net like architecture to estimate the deformable offsets that are later applied to deform a regular convolution and produce spatially-aligned features. Inspired by~\cite{deng2020spatio}, STOPN predicts spatio-temporal offsets that are different at each spatial and temporal position. Moreover, as stated by~\cite{rota2022video}, they apply deformable alignment at feature level to increase alignment accuracy. The third module contains two groups of standard and transposed convolutions to progressively perform feature fusion and upscaling. The input target frame is then $\times4$ upscaled using bicubic interpolation and finally added to the network output to produce the final result. They only process the luma channel (Y) of the input frames because it contains the most relevant information on the scene details. The final result is obtained using the restored Y channel and the original chroma channels upscaled using bicubic interpolation, followed by a RGB conversion.

They train the model for 250,000 iterations using a batch size equal to 32. Patches with the size of $96\times96$ pixels are used, and data augmentation with random flip is applied. They set the temporal neighborhood $N$ to five, hence they stacke the target frame with the two previous and the two subsequent frames. The learning rate was initially set to $10^{-3}$ for the first 200,000 iterations, then reduced to $10^{-4}$ for the remaining iterations. They use MSE as the loss function and optimize it using the Adam optimizer.

\subsection{SRC-B Team}

\begin{figure*}[h]
\centering
\includegraphics[width=.8\linewidth]{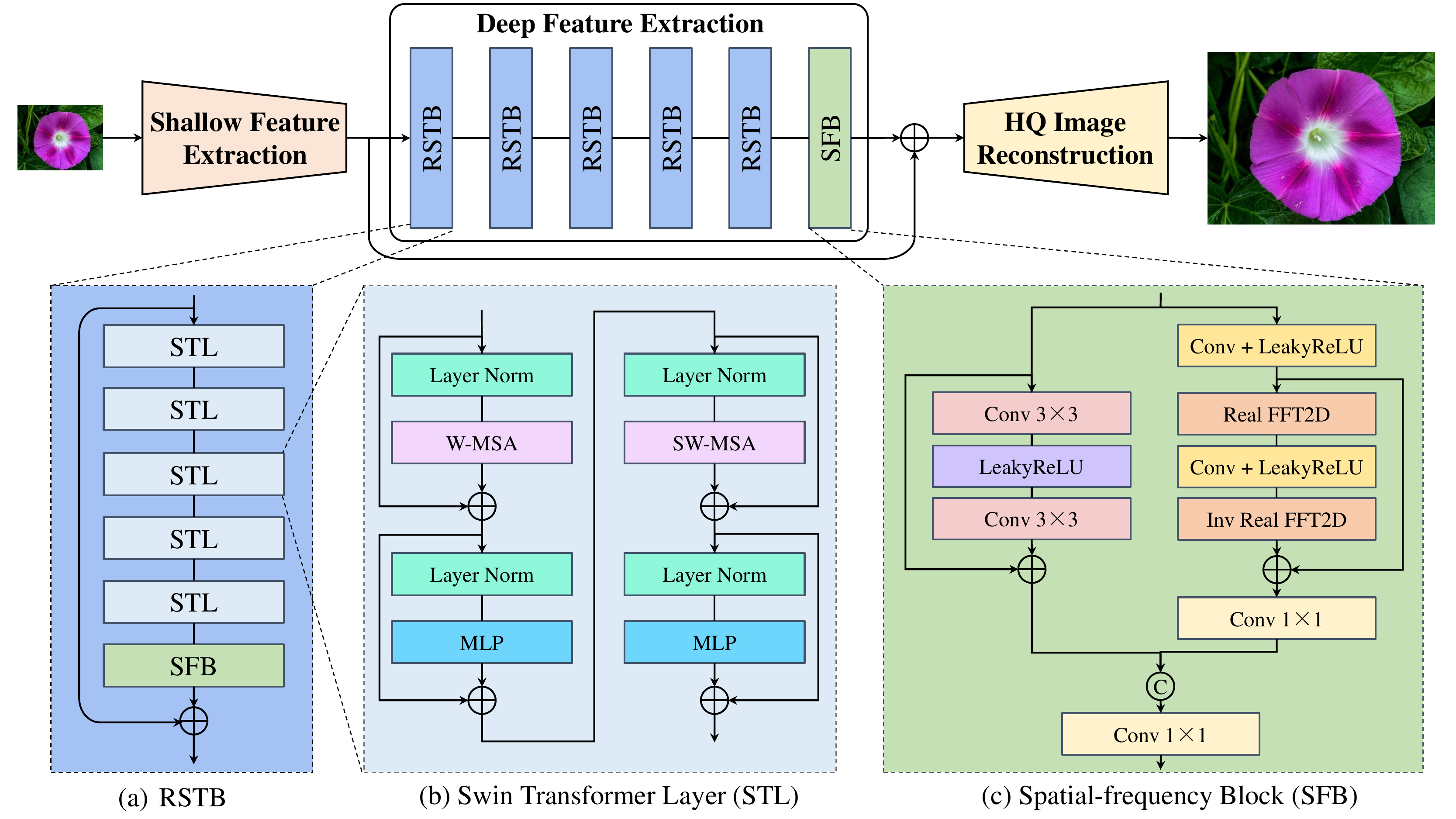}
\caption{The SwinFIR method proposed by the SRC-B Team.}
\label{fig:SwinFIR}
\end{figure*}

Inspired by SwinIR~\cite{liang2021swinir}, the SRC-B Team proposes the SwinFIR method with the Swin Transformer~\cite{liu2021swin} and the Fast Fourier Convolution~\cite{chi2020fast}. As shown in Fig.~\ref{fig:SwinFIR}, SwinFIR consists of three modules: shallow feature extraction, deep feature extraction and high-quality (HQ) image reconstruction modules. The shallow feature extraction and high-quality (HQ) image reconstruction modules adopt the same configuration as SwinIR. The residual Swin Transformerblock (RSTB) is a residual block with Swin Transformerlayers (STL) and convolutional layers in SwinIR. They all have local receptive fields and cannot extract the global information of the input image. The Fast Fourier Convolution has the ability to extract global features, so they replace the convolution (3$\times$3) with Fast Fourier Convolution and a residual module to fuse global and local features, named Spatial-frequency Block (SFB), to improve the representation ability of model.

They use the Adam optimizer with default parameters and the Charbonnier L1 loss~\cite{lai2018fast} to train the model. The initial learning rate is $2\times 10^{-4}$, and they use the cosine annealing learning rate scheduler~\cite{loshchilov2016sgdr} with about 500,000 iterations. The batch size is 32 and patch size is 64. They use horizontal flip, vertical flip, rotation, RGB perm and mixup~\cite{yoo2020rethinking} for data augmentation.

\subsection{USTC-IR Team~\cite{li2022hst}}

\begin{figure}[!t]
    \centering
    \includegraphics[width=.8\textwidth]{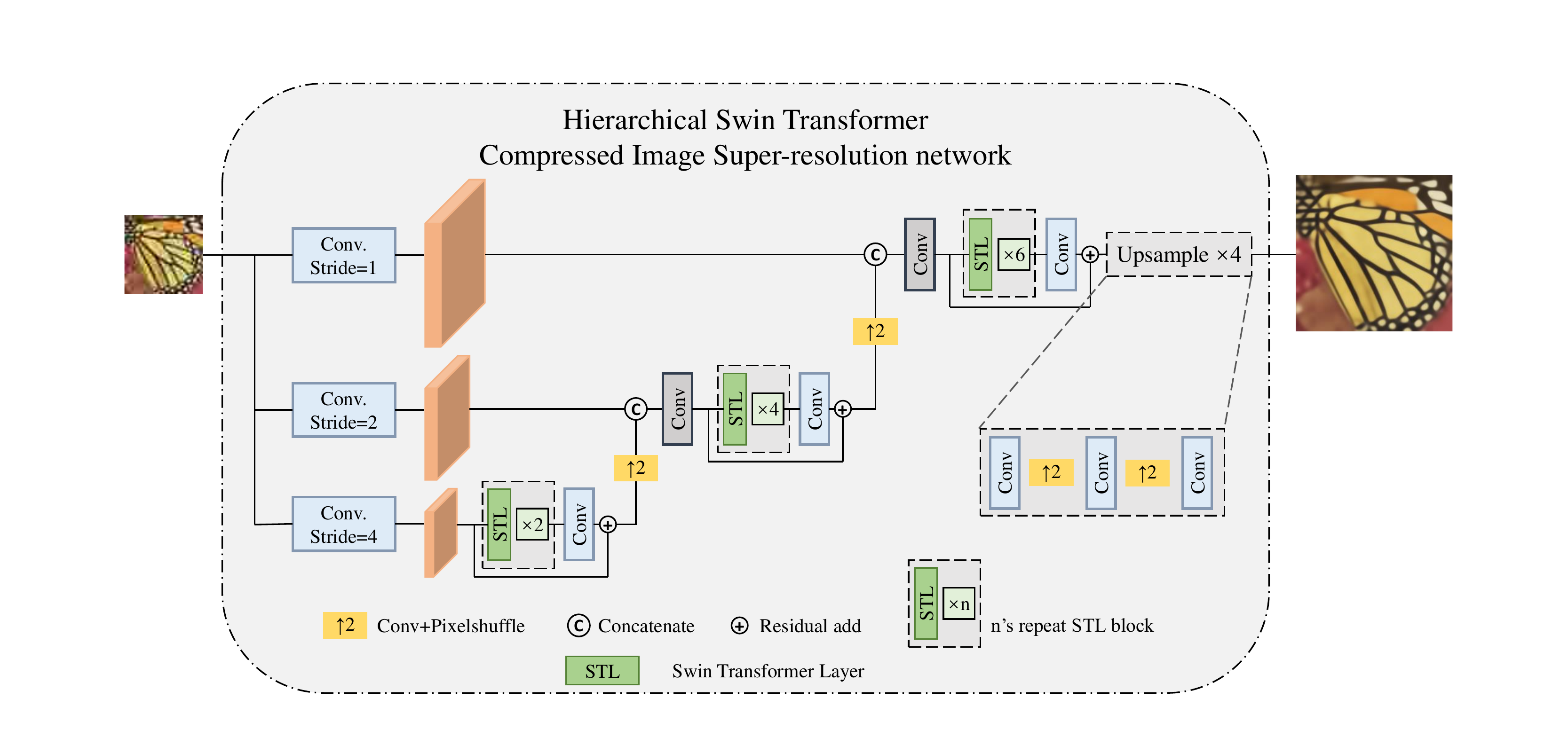}
    \caption{Overview of HST method proposed by the USTC-IR Team. STL block is the Swin Transformer layer from SwinIR~\cite{liang2021swinir}.}
    \label{fig:my_label}
\end{figure}

The USTC-IR Team proposes a Hierarchical Swin Transformer (HST) for compressed image super-resolution, which is inspired by multi-scale-based frameworks ~\cite{pang2020fan,li2020multi,papyan2015multi} and transformer-based frameworks~\cite{liang2021swinir,liang2022vrt}. As shown in Fig.~\ref{fig:my_label}, the network is divided into three branches so that it can learn global and local information from different scales. Specifically, the input image is first downsampled to different scales by convolutions. Then, it is fed to different Residual Swin Transformer Blocks (RSTB) from SwinIR~\cite{liang2021swinir} to obtain the restored hierarchical features from each scale. To fuse the features from different scales, they super-resolve the low-scale feature and concatenate it with the higher feature. Finally, there is a pixel-shuffle block to implement the $4\times$ super-resolution of features.

The training images are paired-cropped into $64\times 64$ patches, and augmented by random horizontal flips, vertical flips and rotations. They train the model by the Adam optimizer with the initial learning rate of $2\times 10^{-4}$. The learning rate is decayed by the factor of 0.5 twice, at the 200,000th and the 300,000 steps, respectively. The network is first trained by Charbonnier loss~\cite{lai2018fast} for about 50,000 steps and finetuned by the MSE loss until convergence.

\subsection{MSDRSR Team}

\begin{figure}[!htb]
    \centering
    \includegraphics[width=0.8\linewidth]{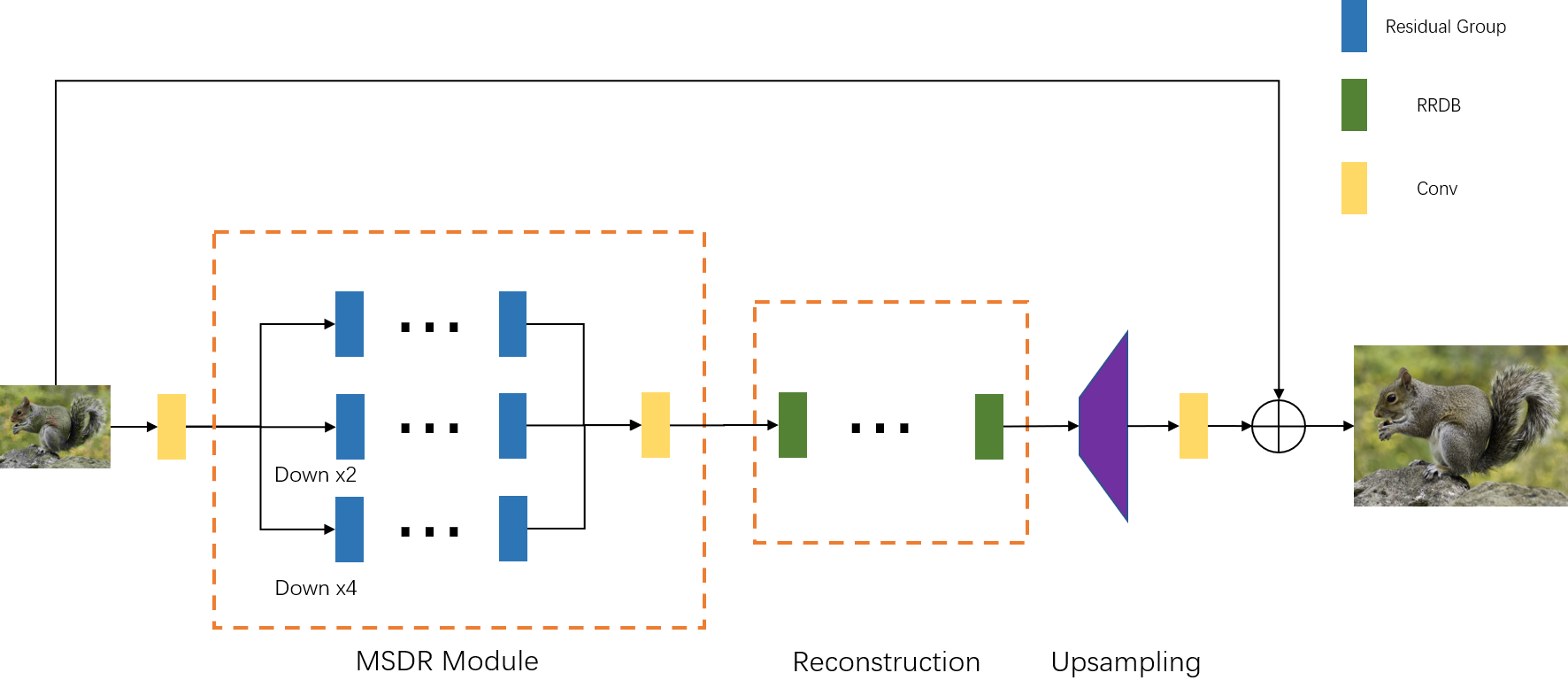}
    \caption{The architecture of the method proposed by the MSDRSR Team. It utilizes a multi-scale degradation removal module, which employs the multi-scale structure to achieve balance between detail enhancement and artifacts removal.}
    \label{fig:framework}
\end{figure}

\begin{figure}[!htb]
    \centering
    \includegraphics[width=0.8\linewidth]{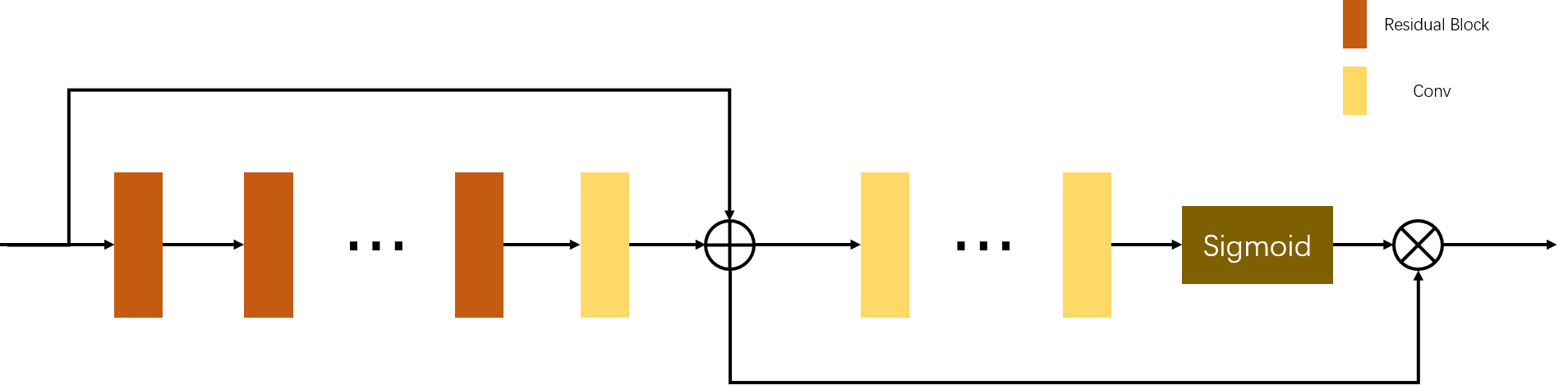}
    \caption{Enhanced Residual Group.}
    \label{fig:erg}
\end{figure}

The architecture of the method proposed by the MSDRSR Team is illustrated in Fig.~\ref{fig:framework}. The details are described in the following.

\textbf{MSDR Module.} In the MSDR Module, a Multi-Scale Degradation Removal (MSDR) module is employed after the first convolutional layer. The MSDR module uses several Enhanced Residual Group (ERG) to extract multi-scale features and can achieve a better trade-off between detail enhancement and compression artifacts removal. The architecture of enhanced residual group (ERG) is illustrated in Fig.~\ref{fig:erg}. ERG removes the channel attention module in each residual block, and adds a high-frequency attention block~\cite{du2022fast} at the end of the residual block. Compared with the original design of residual group, ERG can effectively remove artifacts while reconstructing high-frequency details. Moreover, our proposed ERG is very efficient and does not introduce much runtime overhead.

\textbf{Reconstruction Module.} The reconstruction module is build on ESRGAN~\cite{wang2018esrgan}, which has 23 residual-in-residual dense blocks (RRDB). To further improve the performance~\cite{lin2022revisiting}, they change activation function to SiLU~\cite{elfwing2018sigmoid}.

The MSDRSR employs a two-stage training strategy. In each stage, MSDRSR is first trained with Laplacian pyramid loss with the patch size of 256, and then it is fine-tuned with the MSE loss with the patch size of 640. They augment the training data with random flipping and rotations.
In the first stage, MSDRSR is trained on DF2K~\cite{agustsson2017ntire,timofte2017ntire} for 100,000 iterations with the batch size of 64. It adopts the Adam optimizer with an initial learning rate of $5\times 10^{-4}$. The Cosine scheduler is adopted with the minimal learning rate of $5\times 10^{-5}$. Then, MSDRSR is fine-tuned with learning rate of $10^{-5}$ for 20,000 iterations.
In the second stage, MSDRSR loads pre-trained weights from the first stage, and then they add 10 more randomly initialized blocks to the feature extractor. It is trained on DF2K and DIV8K~\cite{gu2019div8k} datasets. Then MSDRSR adopts the same training strategy as the first stage.

\subsection{Giantpandacv}

Inspired by previous image restoration and JPEG artifacts removal research \cite{FBCNN,chu2022nafssr}, the Giantpandacv Team proposes the Artifact-aware Attention Network (A$^3$Net) that can use the global semantic information of the image to adaptive control the trade-off between artifacts removal and details restored. Specifically, the A$^3$Net uses an encoder to extract image texture features and artifact-aware features simultaneously, and then it adaptively removes image artifacts through a dynamic controller and a decoder. Finally, the A$^3$Net uses some nonlinear-activation-free blocks to build a reconstructor to further recover the lost high-frequency information, resulting in a high resolution image.

\begin{figure}[!t]
	\centering
	\includegraphics[width=0.8\linewidth]{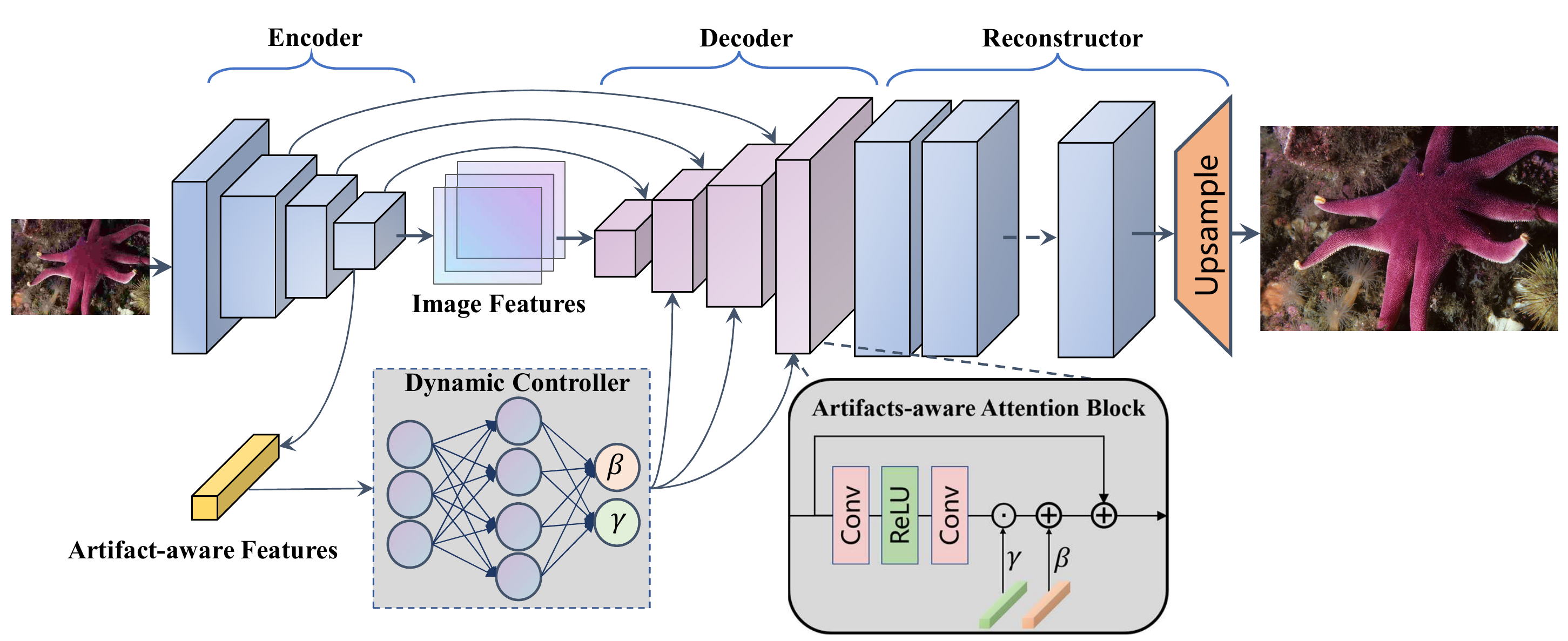}
	\caption{Architecture of the A$^3$Net proposed by the Giantpandacv Team.}
	\label{fig}
\end{figure}

The main architecture of the A$^3$Net is shown in Figure. \ref{fig}, which consists of four components: Encoder, Decoder, Dynamic Controller, and Reconstructor. The details of these modules of A$^3$Net are described as follows. 

\textbf{Encoder:} The Encoder aims to extract the deep features and decouple the latent artifact-aware features from the input image. The Encoder contains four scales, each of which has a skip connection to connect to the decoder. In order to improve the computational efficiency of the network, we use 2 Nonlinear Activation Free (NAF) blocks~\cite{chu2022nafssr} at each scale. The number of output channels in each layer from the first to the fourth scale is set to 64, 128, 256, and 512, respectively. The image features from the encoder are passed into the decoder. At the same time, the global average pooling layer is used to get the artifact-aware features from the image features. 

\textbf{Dynamic Controller:} The dynamic controller is a 3-layer MLP and take as input the artifact-aware features, representing the latent degree of image compression. The main purpose of the dynamic controller is to allow the latent degree of image compression to be flexibly applied to the decoder, thus effectively removing artifacts. Inspired by recent research in spatial feature transform \cite{SFT1,SFT2}, we employ dynamic controller to generate modulation parameters pair ($\gamma, \beta$) which embed on the decoder. Moreover, we used three different final layers of the MLP to accommodate the different scales of features. 

\textbf{Decoder:} The decoder consists of artifact-aware attention blocks with three different scales. The artifact-aware attention blocks mainly removes artifacts by combining image features and embedded artifact-aware parameters ($\gamma, \beta$). The number of artifact-aware attention blocks in each scale is set to 4. It can be expressed as follows:
\begin{equation}
    F_{out} = \gamma \odot F_{in} + \beta
\end{equation}
where $F_{in}$ and $F_{out}$ denote the feature maps before and after the affine transformation, and $\odot$ is referred to as element-wise multiplication.

\textbf{Reconstructor:} The aim of the reconstructor is to  further restore the lost texture details, and then the features are up-sampled to reconstruct a high-resolution image. Specifically, they use a deeper NAF to facilitate the network capturing similar textures over long distances, thus obtaining more texture details. 

\textbf{Implementation and training details:}
The numbers of NAF blocks in the each scale of the Encoder and the Reconstructor are flexible and configurable, which are set to 2 and 8, respectively. For the up-scaling module, they use pixel-shuffle to reconstruct a high-resolution image. During training, the A$^3$Net is trained on the crop training dataset with LR and HR pairs. The input pairs are randomly cropped to $512\times 512$. Random rotation and random horizontal flop are applied for data augmentation. They use the AdamW optimizer with the learning rate of $2\times 10^{-4}$ learning rate to train the model for 1,000,000 iterations and the learning rate is decayed with the cosine strategy. Weight decay is $10^{-4}$ for all the training periodic. 

\subsection{Aselsan Research Team}

\begin{figure}[h!]
\centering
\includegraphics[width=.8\textwidth]{"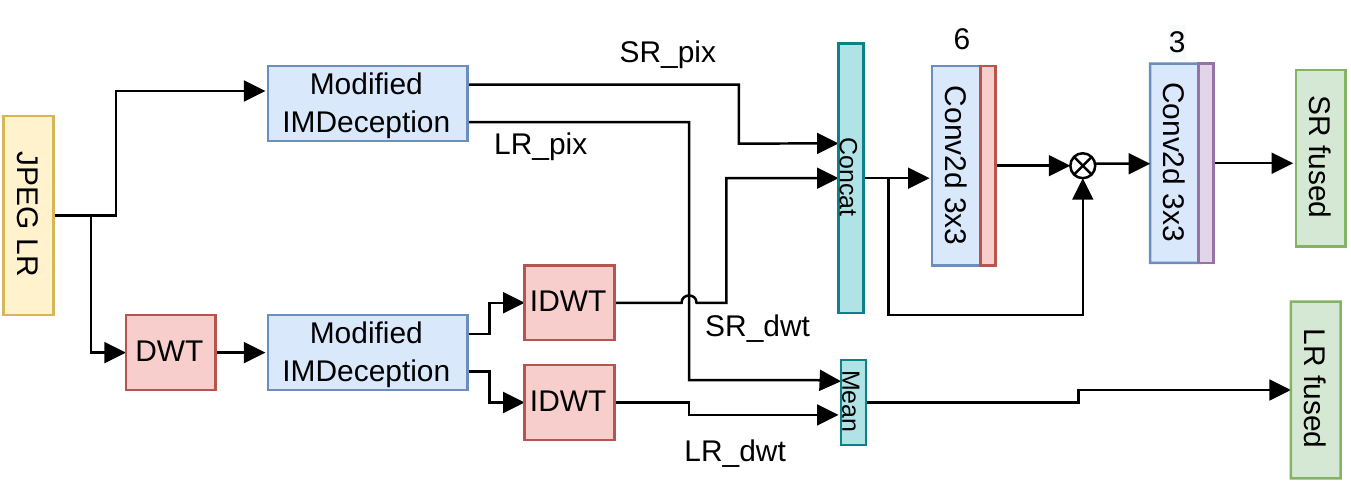"}
\caption{The dual-domain super-resolution network of the Aselsan Research Team.}
\label{"3DIMDeception"}
\end{figure}

The Aselsan Research Team proposes a dual-domain super-resolution network, shown in Fig.~\ref{"3DIMDeception"}. The network utilizes information in both pixel and wavelet domains. The information of these domains are processed in parallel by a modified IMDeception Network~\cite{ayazouglu2022imdeception} to further increase the receptive field and the capability for processing non-local information. The two branches of the modified IMDeception network generates the super-resolved image and the enhanced low-resolution image, respectively. The super-resolved images are fused through a pixel attention network as used in~\cite{bilecen2022efficient} and enhanced low-resolution images are averaged for fusion. These low-resolution outputs are used during training to further guide the network and add the dual capability to the network. To further boost the performance of the entire network, the structure is encapsulated in a geometric ensembling architecture. We used LR\_fused output as well as SR\_fused output for training, using LR as guidance through out the optimization. The loss function we used is as follows
\begin{equation}
L = 0.5\times L2(\text{LR\_fused}-\text{Uncompressed\_LR})+0.5\times L2(\text{SR\_fused}-\text{HR})
\end{equation}
Note that almost entire network is shared for this dual purpose, having a secondary and complementary guidance which is able to boost the performance by around 0.05dB.

They use the Adam optimizer with $\beta_1=0.9, \beta_2=0.999$ for training. The batch size is set to 8, and the training samples are cropped to $512\times 512$. The learning rate is initialized to $5\time 10^{-4}$ and it is decayed with the factor of 0.75 every 200 epochs (800 iteration in each epoch). The model is totally trained for 2,000 epochs.

\subsection{SRMUI Team~\cite{conde2022swin2sr}}

\begin{figure}
    \centering
    \includegraphics[width=.8\textwidth]{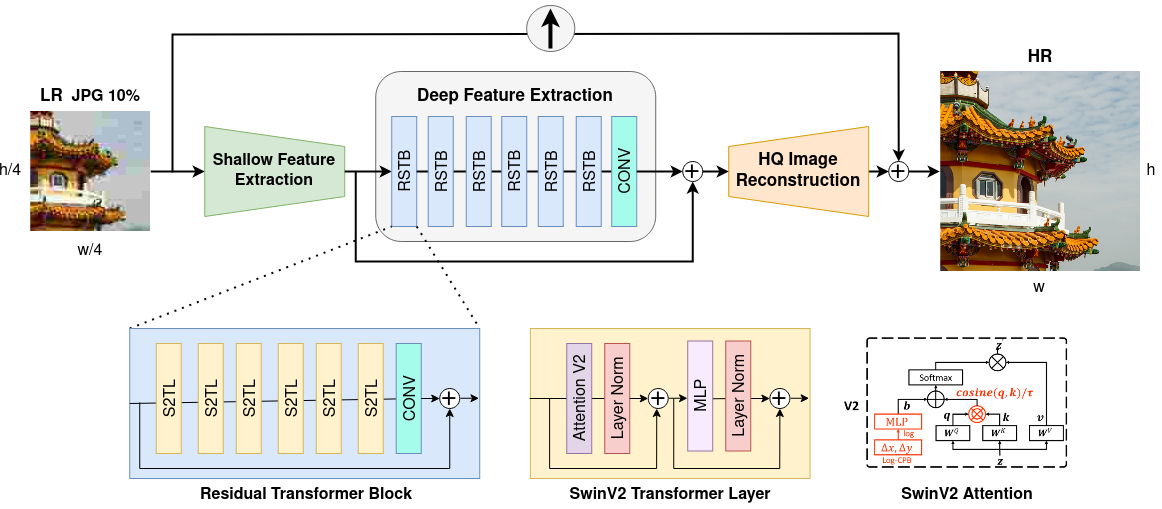}
    \caption{The Swin2SR method of the SRMUI Team.}
    \label{fig:main}
\end{figure}

The method proposed by the SRMUI Team is illustrated in Fig.~\ref{fig:main}. They propose some modifications of SwinIR~\cite{liang2021swinir} (based on Swin Transformer~\cite{liu2021swin}) that enhance the model's capabilities for super-resolution, and in particular, for compressed input SR. They update the original Residual Transformer Block (RSTB) by using the new SwinV2 transformer~\cite{liu2022swin} layers and attention to scale up capacity and resolution.
This method has a classical upscaling branch which uses a bicubic interpolation to recover basic structural information. The output of our model is added to the basic upscaled image to enhance it.
They also explored different loss functions to make the model more robust to JPEG compression artifacts, being able to recover high-frequency details from the compressed LR image, and therefore it is able to achieve better performance.

\subsection{MVideo Team}

\begin{figure}[b]
\centering
\includegraphics[width=.8\textwidth]{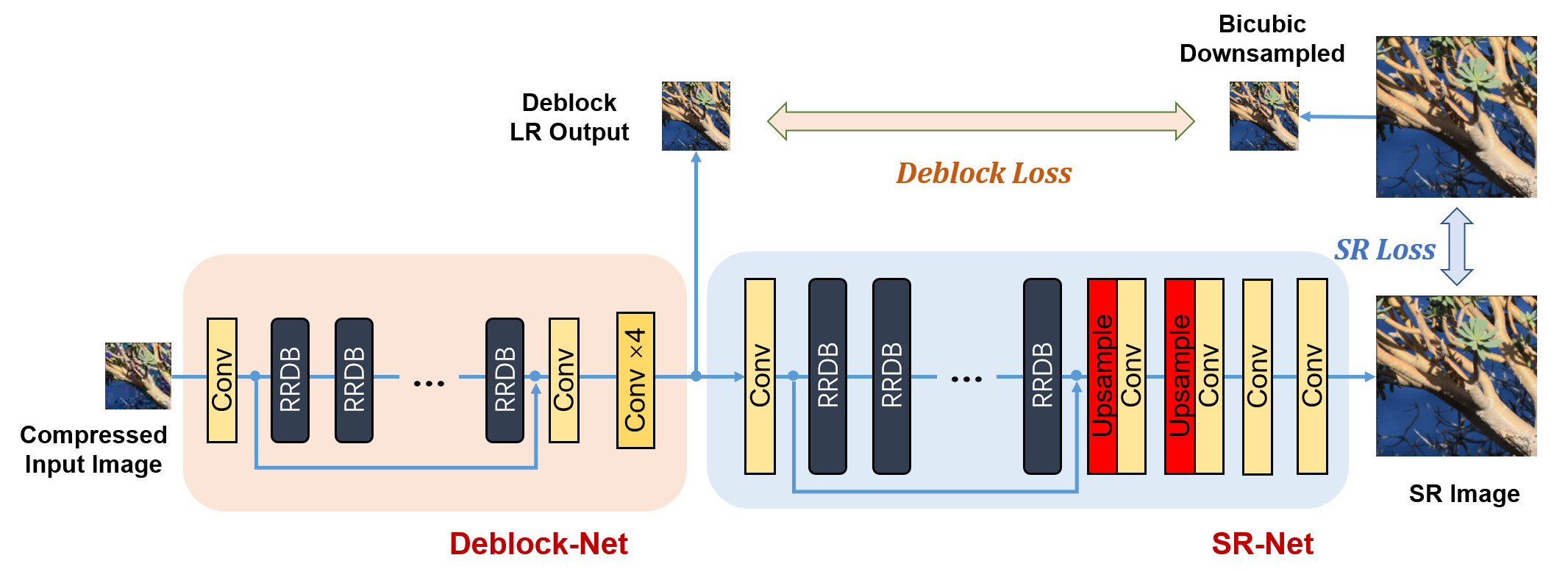}
\caption{The two-stage network proposed by the MVideo Team.}
\label{fig1}
\end{figure}

The MVideo Team proposes a two-stage network for compressed image super resolution. The overall architecture of the network is as Fig.~\ref{fig1}. The Deblock-Net in first stage takes the compressed low resolution image with JPEG blocky artifacts as input and outputs the enhanced low resolution image. Then the SR-Net in the second stage is applied to the enhanced low resolution image, and generate the final SR image. Both networks use RRDBNet~\cite{wang2018esrgan} as implementation, while they remove the pixel unshuffle operation from the first Deblock-Net in the beginning and the upsample operation at last. The SR-Net uses the same hyper-parameters as used in ESRGAN~\cite{wang2018esrgan}.

Based on this pipeline, they train the two networks separately to reduce training time consumption. Firstly, they use the pretrained weights from the official ESRGAN, and load the SR-Net with it. Then the SR-Net is freezed and the deblock loss between bicubic downsampled LR from ground truth HR and deblock output is applied in order to train the Deblock-Net. After the training of the deblock net finishes, they use both deblock loss and SR loss (final output and ground truth) to train the model, with the weight of the deblock loss of 0.01 and the weight of the SR loss of 1.0. Then the model is finetuned only using the SR loss to improve the PSNR of the final output. Detailed training settings are listed below:

\begin{itemize}
    
\item (I) Pre-train the Deblock-Net. Firstly load the pretrained RRDBNet to SR-Net, and then use only deblock loss to train the Deblock-Net. The patch size is 128 and batch size is 32. Training is for 50k iterations using Adam optimizer, and learning rate is initiated with $5\times 10^{-4}$, which decreases with the factor of 0.5 at the 30,000th and 40,000th iterations, respectively.
	
\item (II) End-to-end training of the two-stage network. Using the pretrained weights from stage (I), they train the full network using both the deblock loss (weight=0.01) and the SR loss (weight=1.0), which mainly focuses on the learning of SR-Net. In this stage, patch size, batch size, and Adam optimizer stay the same as above, and they use the CosineAnnealingRestartLR scheduler with all periods of 50,000 for 200,000 iterations.
	
\item (III) Finetuning the last weights using patch size of 512 and batch size of 8. The initial learning rate is set as $2\times 10^{-5}$ and the learning rate decays by the factor at the 20,000th, 30,000th, and 40,000th iterations, respectively. This stage takes totally 50,000 iterations.
	
\item (IV) The last stage finetunes the model from the previous stage for 50,000 iterations, with the patch size of 256 and the batch size of 8, initial learning rate of $5\times 10^{-6}$ and multistep scheduler which decreases the learning rate by the factor of 0.5 at the 20,000th, 30,000th and 40,000th iterations, respectively. The final model is used in the inference phase.

\end{itemize}

\subsection{UESTC+XJU CV Team}

\begin{figure*}[h]
  \centering
  \includegraphics[width=.8\textwidth]{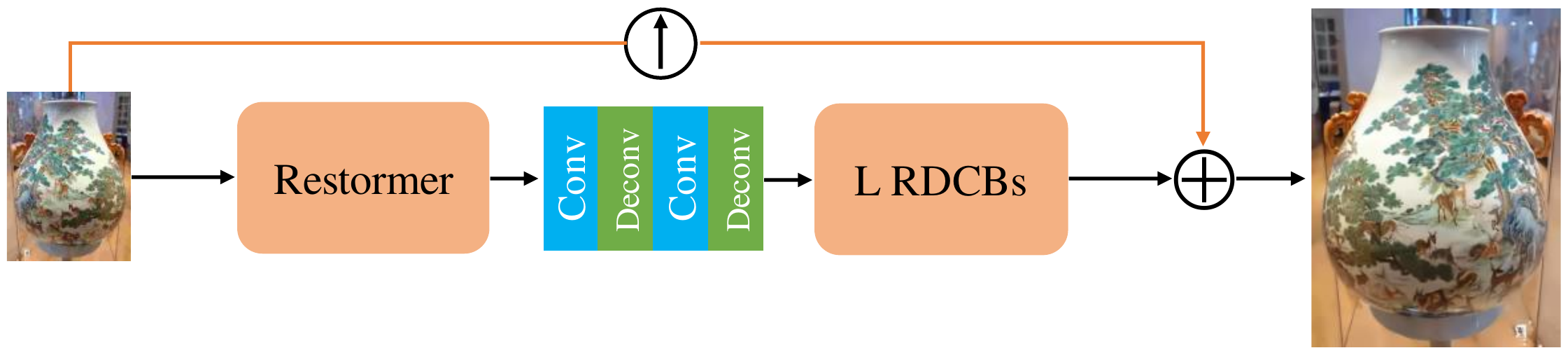}
  \caption{Method of the UESTC+XJU CV Team: based on Restormer \cite{zamir2022restormer}. RDCB: A \textbf{C}hannel-attention \textbf{B}lock is inserted into the \textbf{R}esidual \textbf{D}ense connection block.}
  \label{method}
\end{figure*}

The UESTC+XJU CV Team utilizes Restomer~\cite{zamir2022restormer} for compressed image enhancement. The overall structure is shown in Fig. \ref{method}. First, the compressed image is input into the Restomer (the last layer of the original network is removed), and then upsampled by two convolutional layers and deconvolutional layers. Finally, the feature map is input into the common CNN network, which consists of $L$ (4 in this model) Residual Dense Channel-attention Blocks (RDCBs), and then it is added with the original upsampled image to obtain the reconstructed image. Specially, the channel-attention layer~\cite{zhang2018image} is inserted after the four dense connection layers in the residual dense block~\cite{zhang2018residual}.

In the training process, the raw image is cropped into patches with the size of 256$\times$256 as the training samples,
and the batch size is set to 8.
They also adopt flip and rotation as data augmentation strategies to further expand the dataset.
The model is trained by Adam optimizer~\cite{kingma2014adam} and cosine annealing learning rate scheduler for $3 \times 10^{5}$ iterations.
The learning rate is initially set to $3 \times 10^{-4}$.
They use L2 loss as the loss function.

\subsection{cvlab Team}

\begin{figure*}[t]
\centering
	\includegraphics[width=.7\textwidth]{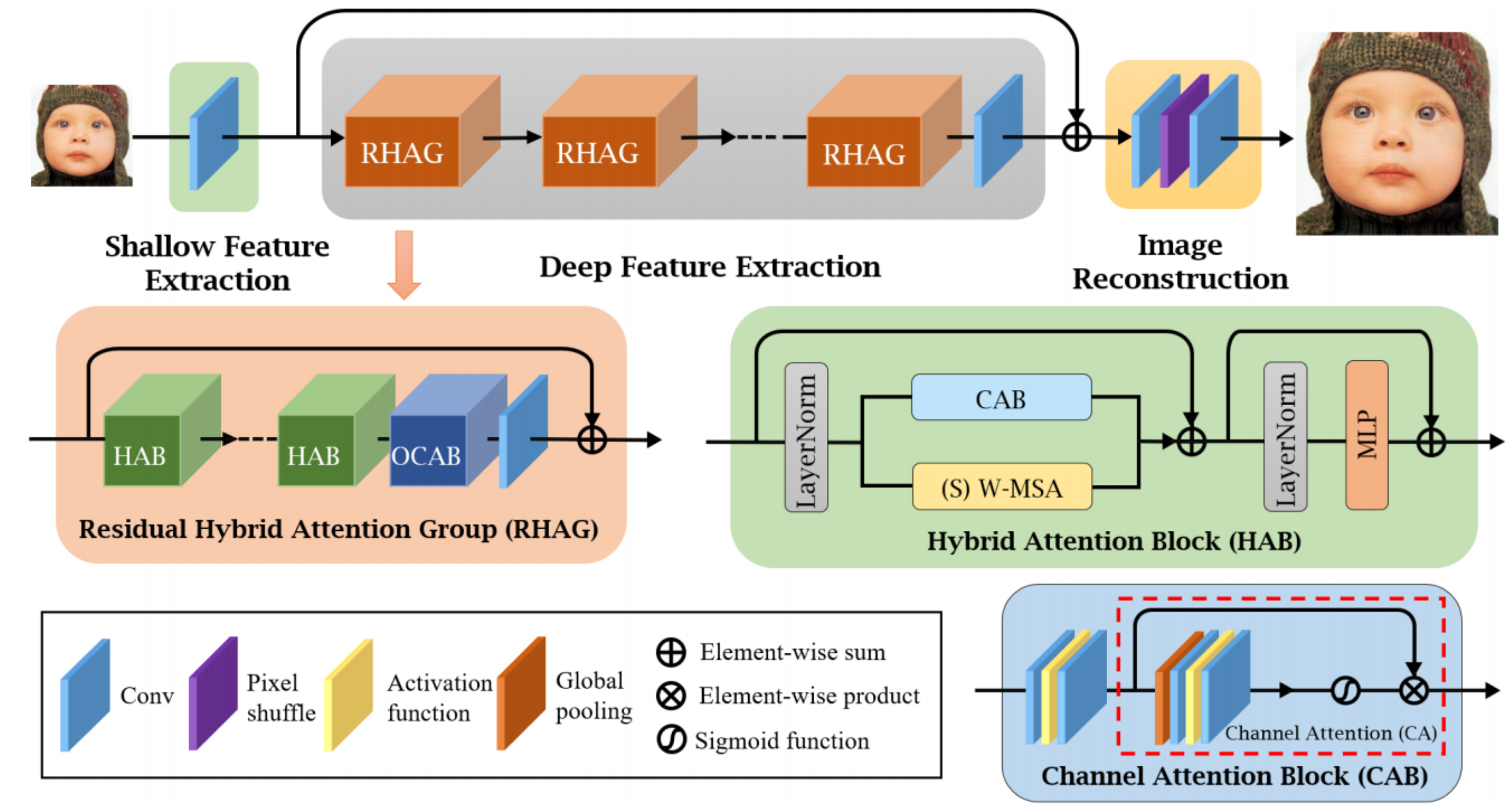}
	\caption{The HAT~\cite{chen2022activating} method used by the cvlab Team in Track 1.}
	\label{fig1}
\end{figure*}

The cvlab Team uses HAT~\cite{chen2022activating} as the solution in Track 1. As show in Fig.~\ref{fig1}, The number of RHAG blocks is set to 6. The number of HAB blocks in each RHAG block is 6. The number of feature channels is 180. In the training process, the raw and compressed sequences are cropped into $64\times 64$ patches as the training pairs, and the batch size is set to 4. They also adopt flip and rotation as data augmentation strategies to further expand the dataset. We train all models using Adam~\cite{kingma2014adam} optimizer with $\beta_{1}=0.9$, $\beta_{2}=0.999$, $\epsilon=10^{-6}$, and the learning rate is initially set to $2\times10^{-4}$ and decays linearly to $5\times10^{-5}$ after 200,000 iterations, which keeps unchanged until 400,000 iterations. Then, the learning rate is further decayed to $2\times10^{-5}$ and $1\times10^{-5}$ until converged. The total number of iterations is 1,000,000. They use L1 loss as the loss function.

\section*{Acknowledgments}
We thank the sponsors of the AIM and Mobile AI 2022 workshops and challenges: AI Witchlabs, MediaTek, Huawei, Reality Labs, OPPO, Synaptics, Raspberry Pi, ETH Z\"urich (Computer Vision Lab) and University of W\"urzburg (Computer Vision Lab).

\section*{Appendix: Teams and affiliations}

\subsection*{AIM 2022 Team}

\noindent\textbf{Challenge:}  

\noindent AIM 2022 Challenge on Super-Resolution of Compressed Image and Video

\noindent\textbf{Organizer(s):} 

\noindent Ren Yang$^{1}$ ({\texttt{ren.yang@vision.ee.ethz.ch}}), 

\noindent Radu Timofte$^{1,2}$ ({\texttt{radu.timofte@uni-wuerzburg.ch}})

\noindent\textbf{Affiliation(s):}

\noindent $^1$ Computer Vision Lab, ETH Z\"urich, Switzerland\\
\noindent $^2$ Julius Maximilian University of W\"urzburg, Germany

\subsection*{VUE Team}
\noindent\textbf{Member(s):} 

\noindent Xin Li$^{1}$ (\texttt{lixin41@baidu.com}), Qi Zhang$^{1}$, Lin Zhang$^{2}$, Fanglong Liu$^{1}$, Dongliang He$^{1}$, Fu li$^{1}$, He Zheng$^{1}$, Weihang Yuan$^{1}$

\noindent\textbf{Affiliation(s):}

\noindent $^\text{1 } $ Department of Computer Vision Technology (VIS), Baidu Inc. \\
$^\text{2 } $  Institute of Automation, Chinese Academy of Sciences

\subsection*{NoahTerminalCV Team}
\noindent\textbf{Member(s):} 

\noindent Pavel Ostyakov (\texttt{ostyakov.pavel@huawei.com}), Dmitry Vyal, Magauiya Zhussip, Xueyi Zou, Youliang Yan

\noindent\textbf{Affiliation(s):}

\noindent Noah's Ark Lab, Huawei

\subsection*{BSR Team}
\noindent\textbf{Member(s):} 

\noindent Lei Li (\texttt{lilei.leili@bytedance.com}), Jingzhu Tang, Ming~Chen, Shijie Zhao

\noindent\textbf{Affiliation(s):}

\noindent  Multimedia Lab, ByteDance Inc.

\subsection*{CASIA LCVG Team}
\noindent\textbf{Member(s):} 

\noindent Yu Zhu$^{1}$ (\texttt{zhuyu.cv@gmail.com}), Xiaoran Qin$^{1}$, Chenghua Li$^{1}$, Cong Leng$^{1,2,3}$, Jian Cheng$^{1,2,3}$

\noindent\textbf{Affiliation(s):}

\noindent $^1$ Institute of Automation, Chinese Academy of Sciences, Beijing, China\\
\noindent $^2$ MAICRO, Nanjing, China \\
\noindent $^3$ AiRiA, Nanjing, China \\

\subsection*{IVL Team}
\noindent\textbf{Member(s):} 

\noindent Claudio Rota (\texttt{c.rota30@campus.unimib.it}), Marco Buzzelli, Simone Bianco, Raimondo Schettini

\noindent\textbf{Affiliation(s):}

\noindent University of Milano - Bicocca, Italy

\subsection*{Samsung Research China – Beijing (SRC-B)}
\noindent\textbf{Member(s):} 

\noindent Dafeng Zhang (\texttt{dfeng.zhang@samsung.com}), Feiyu Huang, Shizhuo Liu, Xiaobing Wang, Zhezhu Jin

\noindent\textbf{Affiliation(s):}

\noindent Samsung Research China – Beijing (SRC-B), China

\subsection*{USTC-IR}
\noindent\textbf{Member(s):} 

\noindent Bingchen Li (\texttt{lbc31415926@mail.ustc.edu.cn}), Xin Li

\noindent\textbf{Affiliation(s):}

\noindent University of Science and Technology of China, Hefei, China\\

\subsection*{MSDRSR}
\noindent\textbf{Member(s):} 

\noindent Mingxi~Li (\texttt{li\_mx\_0318@163.com}), Ding~Liu$^{1}$

\noindent\textbf{Affiliation(s):}

\noindent $^{1}$ ByteDance Inc.\\

\subsection*{Giantpandacv Team}
\noindent\textbf{Member(s):} 

\noindent Wenbin~Zou$^{1,4}$ (\texttt{alexzou14@foxmail.com}), Peijie~Dong$^{2}$, Tian Ye$^{3}$, Yunchen Zhang$^{5}$, Ming Tan$^{4}$, Xin Niu$^{2}$\

\noindent\textbf{Affiliation(s):}

\noindent $^1$ South China University of Technology, Guangzhou, China\\
\noindent $^2$ National University of Defense Technology, Changsha, China \\
\noindent $^3$ Jimei University, Xiamen, China \\
\noindent $^4$ Fujian Normal University, Fuzhou, China \\
\noindent $^5$ China Design Group Inc., Nanjing, China \\

\subsection*{Aselsan Research Team}
\noindent\textbf{Member(s):} 

\noindent Mustafa Ayazoğlu (\texttt{mayazoglu@aselsan.com.tr})

\noindent\textbf{Affiliation(s):}

\noindent Aselsan (\hyperlink{www.aselsan.com.tr}{www.aselsan.com.tr}), Ankara, Turkey \\

\subsection*{SRMUI Team}
\noindent\textbf{Member(s):} 

\noindent Marcos V. Conde$^1$ (\texttt{marcos.conde-osorio@uni-wuerzburg.de}), Ui-Jin Choi$^2$, Radu Timofte$^1$

\noindent\textbf{Affiliation(s):}

\noindent $^1$ Computer Vision Lab, Julius Maximilian University of W\"urzburg, Germany
\noindent $^2$ MegaStudyEdu, South Korea

\subsection*{MVideo Team}
\noindent\textbf{Member(s):} 

\noindent Zhuang Jia (\texttt{jiazhuang@xiaomi.com}), Tianyu Xu, Yijian Zhang

\noindent\textbf{Affiliation(s):}

\noindent Xiaomi Inc.

\subsection*{UESTC+XJU CV Team}
\noindent\textbf{Member(s):} 

\noindent Mao Ye (\texttt{cvlab.uestc@gmail.com}), Dengyan Luo, Xiaofeng Pan

\noindent\textbf{Affiliation(s):}

\noindent University of Electronic Science and Technology of China, Chengdu, China

\subsection*{cvlab Team}
\noindent\textbf{Member(s):} 

\noindent Liuhan Peng$^1$ (\texttt{pengliuhan@gmail.com}), Mao Ye$^2$

\noindent\textbf{Affiliation(s):}

\noindent $^1$ Xinjiang University, Xinjiang, China

\noindent $^2$ University of Electronic Science and Technology of China, Chengdu, China

%
%
\bibliographystyle{splncs04}
\bibliography{egbib}
\end{document}